\documentclass{article}

\usepackage[english]{babel}

\usepackage[letterpaper,top=2cm,bottom=2cm,left=3cm,right=3cm,marginparwidth=1.75cm]{geometry}

\usepackage{amsmath, amsthm, amssymb}
\usepackage{mathtools}

\usepackage{graphicx}
\usepackage{subcaption}
\usepackage{tikz}

\usepackage[ruled, vlined]{algorithm2e}

\usepackage{booktabs}
\usepackage[table]{xcolor}

\usepackage{enumitem}

\usepackage[colorlinks=true, allcolors=blue]{hyperref}
\usepackage{cleveref}

\DeclareMathOperator{\ar}{ar}
\DeclareMathOperator{\vars}{vars}
\DeclareMathOperator{\atoms}{atoms}
\DeclareMathOperator{\free}{free}

\DeclareMathOperator{\anc}{anc}

\newcommand{\ith}{^{\text{th}}}
\newcommand{\bvars}{\textbf{vars}}
\newcommand{\bdom}{\textbf{dom}}

\newcommand{\precbar}{\mathrel{\overline{\prec}}}

\newcommand{\vroot}{v_\text{root}}

\newcommand{\bucket}[2]{\texttt{bucket}_{#1}(#2)}
\newcommand{\ext}{\text{ext}}

\newtheorem{theorem}{Theorem}[section]
\newtheorem{definition}[theorem]{Definition}
\newtheorem{lemma}[theorem]{Lemma}

\newtheorem{proposition}[theorem]{Proposition}
\newtheorem{example}[theorem]{Example}

\newtheorem{claim}[theorem]{Claim}

\AddToHook{env/proposition/begin}{\crefalias{theorem}{proposition}}
\AddToHook{env/lemma/begin}{\crefalias{theorem}{lemma}}
\AddToHook{env/example/begin}{\crefalias{theorem}{example}}
\AddToHook{env/definition/begin}{\crefalias{theorem}{definition}}
\AddToHook{env/corollary/begin}{\crefalias{theorem}{corollary}}
\AddToHook{env/claim/begin}{\crefalias{theorem}{claim}}
\AddToHook{env/remark/begin}{\crefalias{theorem}{remark}}
\AddToHook{env/conjecture/begin}{\crefalias{theorem}{conjecture}}

\usepackage{authblk}

\title{Order-Dependent Complexity of Enumeration and Direct Access for Conjunctive Queries on Dynamic Data}
\author[1,2]{Antoine Amarilli}
\author[3]{Nofar Carmeli}
\author[3]{Clémence Dumoulin}
\affil[1]{LTCI, Télécom Paris, Institut polytechnique de Paris, France}
\affil[2]{Univ.\ Lille, Inria, CNRS, Centrale Lille, UMR 9189 CRIStAL, France}
\affil[3]{Inria, LIRMM, Univ.\ Montpellier, CNRS, France}
\date{}

\begin{document}
\maketitle

\begin{abstract}
This note studies the fine-grained complexity of ranked enumeration and ranked direct access of conjunctive query answers over a dynamic database. 
We focus on \emph{lexicographically} sorted answers, defined by an ordering on the query's free variables. 

From the literature, when no particular order is required, we know that when the query is self-join-free and q-hierarchical, we can build a data structure for enumeration (resp.\ direct access) with linear (resp.\ linearithmic) preprocessing and constant (resp.\ logarithmic) update time in the size of the database. For enumeration (resp.\ direct access), this data structure yields an algorithm that enumerates all query answers (resp.\ that returns the $k\ith$ query answer) with constant delay (resp.\ logarithmic access time).
Our goal is to extend these results to ranked enumeration and ranked direct access, with the ranking function defined by a lexicographic order.

We define a property on the order over the variables of $Q$, which we call being \emph{hierarchy-consistent}.
Given a user-specified query and \emph{hierarchy-consistent} order, we first propose, for ranked enumeration under updates, an algorithm with linear preprocessing and logarithmic update time
that enumerates all the query answers according to the user-specified order with constant delay. 
Moreover, for ranked direct access under updates, we propose an algorithm with linear preprocessing and logarithmic update time that returns the $k\ith$ tuple according to the user-specified order with logarithmic access time.

In contrast, given a user-specified query and \emph{non-hierarchy-consistent} order, we show a lower bound for ranked direct access under updates.
\end{abstract}

\section{Introduction}

For a conjunctive query (or CQ) $Q$, given a database $D$, materializing the answers $Q(D)$ can be costly: the number of answers $|Q(D)|$ can be up to $|D|^{|Q|}$. Producing or holding the full answer set at once is therefore often impractical.
Instead, it is more natural to find approaches that allow us to access partial
information about the query result without materializing it.
In particular, we may want to access 
the $k\ith$ answer to $Q$ on~$D$, or to enumerate the answers one by one without
repetitions.

These tasks are typically formalized in database theory in two ways. The first
one is \emph{enumeration}: rather than producing all tuples at once, an
enumeration algorithm outputs them one by one, with a guarantee of bounded delay
between two consecutive tuples. The second one is \emph{direct access}: this
is the task of outputting, on an input $k$, the $k\ith$ answer for $Q$ over $D$.

It is well-known from the work of Bagan, Durand, and
Grandjean~\cite{bagan2007acyclic} that we can enumerate the answers to
self-join-free CQs with linear preprocessing and constant delay in data complexity
provided that these queries satisfy the so-called \emph{acyclic free-connex
condition}; this restriction is (conditionally) necessary within the class of 
self-join-free CQs, and a characterization of tractable CQs with self-joins is
still open~\cite{carmeli2023conjunctive,rouvroy2026constant}. What is more,
Brault-Baron~\cite{brault2013pertinence} has shown that \emph{direct access} was
possible with linearithmic preprocessing and logarithmic access time in data complexity.
We also note the line of work on \emph{factorized databases}~\cite{Olteanu2016FactorizedDatabase}, which represent query results compactly using a structure derived from the query, and enumerate the answers with constant delay when no particular order is required or when the user-defined order is compatible with this structure.

The literature has then extended these questions along two independent axes: supporting efficient \emph{updates} to the underlying database (and updating the structure used for enumeration or direct access without recomputing it from scratch); and providing results in a specific \emph{order} (e.g., in some lexicographic order specified by the user). The goal of the present work is to bridge these two directions. We present existing results on each of them in turn.

First, in the dynamic setting, \cite{Berkholz2017AnsweringCQ} show that q-hierarchical conjunctive queries can be enumerated with constant delay and constant update time after linear preprocessing, and that for self-join-free CQs this classification is (conditionally) optimal. An  
analogous result for direct access is shown in~\cite{Keppeler2020Answering}.
Further works have investigated which guarantees could be provided for CQs beyond this class; this is not the focus of the present work, but we review these directions nevertheless. 
Dynamic Yannakakis~\cite{Idris2017DynamicYannakakis} maintains constant-delay
enumeration for all free-connex queries (i.e., more generally than
q-hierarchical conjunctive queries)
with a more costly update time (namely, it is linear in the input database).
This was then refined in \cite{Wang2023WithoutJoins} with an update cost that depends on the update sequence, which becomes constant for insertion-only sequences.

We now review the different types of data structures that are used to show efficient enumeration and direct access guarantees on dynamic data; we will re-use some of them in the present work. \cite{Keppeler2020Answering} use a circuit-like structure that stores the assignments of each variable of the \mbox{q-tree} in AVL trees. Circuits could also describe such a structure, but they are not so convenient to maintain in the dynamic setting, given that the domain of assignments is not fixed and updates add or remove edges of the circuit. In contrast, \cite{Kara2024TractableCQ} use view-trees, whose interface and implementation can be defined separately; we re-use the structure of view-trees in the present paper.
In all these works, the answers are produced in an arbitrary order determined by the data structure, and ranked enumeration or ranked direct access under updates is not handled.

Second, for ranked enumeration or direct access
in the static setting,
where answers are output according to a ranking function, an important earlier work is \cite{tziavelis2025any}. For \emph{acyclic free-connex queries}, they build a data structure in linearithmic data complexity, after which the answers to~$Q$ on~$D$ can be enumerated in any lexicographic order with logarithmic delay between consecutive tuples. \emph{Ranked} direct access, which returns the $k$-th answer in the user-defined order, has been studied by \cite{Carmeli2023}, who also build a data structure in quasilinear time, with logarithmic access time. Circuits, which are compact representations of the answers, have also been used for direct access, for conjunctive queries with negation~\cite{capelli2026dacqnlong}, and for enumeration beyond conjunctive queries through knowledge compilation~\cite{Hu2026}.
None of these works considers updates to the database.

Hence, our contribution sits at the intersection of these two lines of work: which orders can we support in the dynamic setting? Since we consider the dynamic setting, we build on the dichotomy established by \cite{Keppeler2020Answering} (and \cite{Berkholz2017AnsweringCQ} for enumeration), which identifies \emph{q-hierarchical} queries as the class for which linear preprocessing, constant delay or logarithmic access, and constant updates are possible under updates. We therefore restrict our attention to this class of queries, and we maintain the answers with view-trees~\cite{Kara2024TractableCQ}.

\paragraph{Contributions.}
This note presents the preliminary results that we have obtained for ranked enumeration and ranked direct access to conjunctive queries under updates. We focus on q-hierarchical queries, because otherwise intractability holds (for self-join-free queries, and conditionally) even in the absence of ranking functions. We study lexicographic orders defined by an order on the query's free variables, and introduce a condition on such orders, called being \emph{hierarchy-consistent}. We show that, for hierarchy-consistent orders, we can perform ranked enumeration and ranked direct access following a lexicographic ordering, with the same favorable guarantees as in the case without a specific ordering, up to a logarithmic update time instead of constant update time for enumeration. By contrast, we show that, for non-hierarchy-consistent orders, a conditional intractability result rules out efficient ranked direct access. 

\paragraph{Paper structure.}
The paper is organized as follows: \Cref{sec:prelim} introduces the notations and concepts necessary to understand the paper. In \Cref{sec:data_structure}, we define the data structure used for ranked enumeration and ranked direct access in the dynamic setting, and give an algorithm to build it.
In \Cref{sec:HC}, we define a property on the user-defined order, called hierarchy-consistency, and give two algorithms for ranked enumeration and ranked direct access in the dynamic setting when the order satisfies this property, with bounds in combined complexity. In \Cref{sec:NonHC}, we show that when the order does not satisfy this property, ranked direct access in the dynamic setting is conditionally not achievable, in data complexity, with polynomial preprocessing, sublinear access time, and sublinear update time.

\section{Preliminaries}\label{sec:prelim}

\subsection{Basic Notions}

\paragraph{Database.} 
A \emph{schema} $\sigma$ is a finite set of relation symbols $\{R_1,\dots,R_s\}$, with every $R\in\sigma$ having a fixed arity $\ar(R)$. 
A \emph{database instance} $D$ is of the form $D = \{R_1^D,\dots,R_s^D\}$ where, for each $R \in \sigma$, $R^D \subseteq \bdom^{\ar(R)}$ is a finite set of tuples $(a_1, \dots, a_{ar(R)})$, and $\bdom$ is a set of constant values called the \emph{domain}, on which we assume a total order.
The size $|D|$ of the database instance $D$ is the number of tuples that it stores, i.e. it is $|D| \coloneq \sum_{R\in \sigma}|R^D|$. 

We fix an infinite set of \emph{variables} denoted by \bvars. For a tuple $t = (a_1,\ldots,a_l)$ indexed by variables $(x_1,\ldots,x_l) \in \bvars^l$ and a set $X \subseteq \{x_1,\ldots,x_l\}$, we write $t[X]$ for the \emph{restriction} of $t$ to the variables of $X$, preserving the order of $t$. Conversely, for a set of variables $Y \supseteq \{x_1,\ldots,x_l\}$ ordered by a total order $\prec$ that agrees with $(x_1,\ldots,x_l)$ on
$\{x_1,\ldots,x_l\}$, a tuple $t'$ indexed by $Y$ is an \emph{extension} of $t$ if $t'[\{x_1,\ldots,x_l\}] = t$, preserving the order $\prec$.

\paragraph{Updates.} 
We consider two types of updates: insertions and deletions. 
An \emph{insertion} is defined as a command $\texttt{insert}~R(a_1,\dots,a_{\ar(R)})$. Applied to a database instance $D$, this results in a database instance $D'$ with $ R^{D'} \coloneq R^D \cup \{(a_1,\dots,a_{ar(R)})\}$. 
Similarly, a \emph{deletion} is defined as a command $\texttt{delete}~R(a_1,\dots,a_{\ar(R)})$. Applied to a database instance $D$, this results in a database instance $D'$ with $ R^{D'} \coloneq R^D \setminus{\{(a_1,\dots,a_{\text{ar}(R)})\}}$.
Note that \texttt{insert} does not change $D$ if the inserted tuple already exists in $D$; likewise, $D$ is unchanged when we delete a tuple that $D$ does not contain.

\paragraph{Queries.}
A \emph{conjunctive query} (CQ) is defined as $Q(\bar{x}) \coloneq \exists \bar{y} (\psi_1 \land \cdots \land \psi_d)$ where $\bar{x} \coloneq (x_1, \ldots, x_l)$, $\bar{y} \coloneq (y_1, \ldots, y_k)$ and each $\psi_i$ is an \emph{atom} of the form $R(z_1, \ldots,z_{\ar(R)})$. We mainly focus on CQs that are \emph{self-join-free}, written SJFCQ, if no relation symbol occurs more than once in the query.
For an atom $\psi$ or a query $Q$, we write $\vars(\psi) \subseteq \bvars$ and $\vars(Q) \subseteq \bvars$ for the set of variables occurring in $\psi$ and $Q$, respectively. For a variable $x \in \vars(Q)$, we denote by $\atoms_Q(x)$ the finite set of atoms in $Q$ in which $x$ appears, i.e. $\atoms_Q(x) = \{\psi \in Q \mid x \in \vars(\psi)\}$.
Moreover, the set of variables $\vars(Q)$ is partitioned into \emph{free} variables $\free(Q) \coloneq \{x_1,\ldots,x_l\}$ and \emph{quantified} variables $\{y_1, \ldots, y_k\}$.
In addition, for a CQ $Q(\bar{x}) \coloneq \exists \bar{y} (\psi_1 \land \cdots \land \psi_d)$ and a set of variables $X \subseteq \free(Q)$, the \emph{projection} of $Q$ over $X$, denoted by $\Pi_X(Q)$, is the CQ defined as $\Pi_X(Q) \coloneq \exists \bar{y} \exists (\bar{x} \setminus X)(\psi_1 \land \cdots \land \psi_d)$. 

Finally, given a query $Q(x_1, \ldots, x_l)$ and a database instance $D$, the set of \emph{answers} to $Q$ over $D$, i.e. the set of tuples resulting from applying $Q$ to $D$, is defined as $Q(D) \coloneq \{(a_1, \ldots, a_l) \in \bdom^l \mid D \models Q(a_1,\ldots,a_l)\}$. Note that $\Pi_X(Q)(D) = \{t[X] \mid t \in Q(D)\}$, i.e., the answers of the projection of a query over $X$ are the restrictions of its answers to $X$.

\subsection{Problem Definition}

The model of computation is the RAM model with uniform cost measure and word size $\Theta(\log n)$. Under this assumption, additions and multiplications on integers of size $O(\log n)$, as well as lookup table and array accesses, take constant time.

Given this computation model, we consider two query evaluation problems throughout this paper: \emph{ranked enumeration under updates} and \emph{ranked direct access under updates}. To define these problems, we first need to introduce the notion of \emph{orders}. 

\paragraph{Order on answers.}

Let $D$ be a database instance, $Q$ a query, and $\prec$ a total order over $\free(Q)$. Let $(x_1,\ldots,x_l)$ be the free variables of $Q$ listed according to this order. We define an \emph{order on answers} $\precbar$ on $Q(D)$ induced by $\prec$, called the \emph{lexicographic order}, as follows. For two answers $t_1,t_2 \in Q(D)$, we say that $t_1 \precbar t_2$ if there exists $j \in [l]$ such that we have $t_1[x_i] = t_2[x_i]$ for all $i < j$ and we have $t_1[x_j] < t_2[x_j]$, where $<$ denotes the total order on the domain values. 

\paragraph{Enumeration.}
The \emph{enumeration} problem consists of two phases. Given a database instance $D$ and a query $Q$, the \emph{preprocessing phase} builds a data structure \texttt{D}.
The \emph{enumeration phase} outputs all tuples in $Q(D)$ without repetition, and finally emits a special value \texttt{end-of-enumeration} to signal the end of enumeration. 
We consider several extensions of this problem. First, if a total order $\prec$ over $\free(Q)$ is given before the preprocessing, the task of \emph{ranked enumeration} asks us to output all tuples in $Q(D)$ in the lexicographic order $\precbar$ on $Q(D)$ induced by $\prec$.
Second, if an update command ($\texttt{insert}~R(a_1, \ldots, a_{\ar(R)})$ or $\texttt{delete}~R(a_1, \ldots, a_{\ar(R)})$) is issued after the preprocessing, letting $D$ be the original database instance and $D'$ be the database instance obtained by applying the update command to~$D$, then the task of \emph{enumeration under updates} asks us to modify the data structure \texttt{D} to a data structure \texttt{D'} that represents the updated database instance, i.e., which allows us to enumerate the tuples of $Q(D')$. If the update command occurs during the enumeration phase, we interrupt enumeration and restart it from scratch on the new database $D'$.
Combining both variants, the \emph{ranked enumeration under updates} is the version where we are given the database instance $D$, the query $Q$, and a total order $\prec$ over $\free(Q)$, and where we must be able to enumerate tuples of $Q(D)$ in the order $\precbar$ and also support updates to~$D$ in the sense above.

\paragraph{Direct access.}
The \emph{direct access} problem consists of two phases. Given a database instance $D$ and a query $Q$, the \emph{preprocessing phase} builds a data structure \texttt{D}. Then, in the \emph{access phase}, we receive user queries consisting of an index $k \geq 0$, and we must return the $k\ith$ tuple of $Q(D)$ in an implementation-defined order, or a special value \texttt{out-of-bounds} if $k \geq |Q(D)|$.
We again extend this problem to support lexicographic orders and to support updates. First, if a total order $\prec$ over $\free(Q)$ is given with $D$ and $Q$ before the preprocessing, then the task of \emph{ranked direct access} asks us to support access queries where the implementation-defined order on tuples must coincide with the lexicographic order $\precbar$ on $Q(D)$, induced by $\prec$.
Second, the task of \emph{ranked direct access under updates} asks us to support update commands ($\texttt{insert}~R(a_1, \ldots, a_{\ar(R)})$ or $\texttt{delete}~R(a_1, \ldots, a_{\ar(R)})$) after the preprocessing, and again we must modify the data structure \texttt{D} so that it represents the updated database instance $D'$ and can be used to access tuples in the new set of answers $Q(D')$. 
Combining both variants, the \emph{ranked direct access under updates} problem is the version where we are given the database instance $D$, the query $Q$, and a total order $\prec$ over $\free(Q)$, and where we must be able to support queries on the lexicographic order induced by~$\prec$ and also to support updates to~$D$ in the sense above.

\paragraph{Complexity measures.}
For the query evaluation problems defined earlier, we measure complexity using \emph{combined complexity}, i.e., as a function of the database size $|D|$ and of the \emph{size} $|Q|$ of the query~$Q$, i.e., the length of $Q$ when written as a word over the alphabet $\sigma \cup\vars(Q) \cup \{\exists, \land, (,)\}$.

For each query evaluation problem, we consider the \emph{preprocessing time} $t_p$, the \emph{update time} $t_u$ when supporting updates, and either the \emph{delay time} $t_d$ for enumeration problems or the \emph{access time} $t_{a}$ for direct access problems.
For ranked enumeration (with or without updates), the \emph{delay time} $t_d$ is the maximum time between any two consecutive outputs, before the first output, and between the last output and the special value \texttt{end-of-enumeration}. For ranked direct access (with or without updates), the \emph{access time} $t_a$ is the time to return the $k\ith$ tuple or the special value \texttt{out-of-bounds}. 
Let $f_p$, $f_d$, $f_a$ and $f_u$ be numeric functions. For ranked enumeration under updates (resp., ranked direct access under updates), an algorithm is said to be in time $\langle f_p, f_d, f_u \rangle$ (resp., $\langle f_p, f_a, f_u \rangle$, if for every database instance $D$ and query $Q$, we have the preprocessing time $t_p = O(f_p(|D|, |Q|))$, the delay time $t_d = O(f_d(|D|, |Q|))$ (resp., the access time $t_a = O(f_a(|D|, |Q|))$) and the update time $t_u = O(f_u(|D|, |Q|))$.

Note that the task of (ranked) direct access (under updates) generalizes the task of (ranked) enumeration (under updates). Indeed, for every numeric function $f_p, f, f_u$, if we have an algorithm in time $\langle f_p, f, f_u \rangle$ for (ranked) direct access under updates, then we have an algorithm in time $\langle f_p, f, f_u \rangle$ for (ranked) enumeration under updates, and likewise without updates. We can perform (ranked) enumeration (under updates) with (ranked) direct access (under updates) as follows. Once the data structure \texttt{D} is built, in $O(f_p(|D|))$, from a q-hierarchical SJFCQ $Q$, a database instance $D$ and whenever it exists, an order $\prec$ over $\free(Q)$, we can return the $k\ith$ answer for $k = 1, 2, 3, \ldots$ in several queries in the access phase, with $O(f(|D|))$ delay. Finally, whenever an update occurs, we modify \texttt{D} in $O(f_u(|D|))$, and restart queries in the access phase with $k = 1$.
    
\subsection{Initial Observations}

The complexity of enumeration and direct access of conjunctive queries on dynamic data was studied in \cite{Keppeler2020Answering}, which introduced the class of q-hierarchical CQs. 

We now recall the definition of this class and the lower bounds and upper bounds
that it yields.

\begin{definition}\label{def:q-hierarchical}
    A CQ $Q$ is \emph{q-hierarchical} if for any two variables $x_1, x_2 \in \vars(Q)$ the following is satisfied:
    \begin{enumerate}[label=(\roman*)]
        \item $\atoms_Q(x_1) \subseteq \atoms_Q(x_2)$ or $\atoms_Q(x_1) \supseteq \atoms_Q(x_2)$ or $\atoms_Q(x_1) \cap \atoms_Q(x_2) = \emptyset$, and
        \item \label{item:qhier-ii} if $\atoms_Q(x_1) \subsetneq \atoms_Q(x_2)$ and $x_1 \in \free(Q)$, then $x_2 \in \free(Q)$.
    \end{enumerate}
\end{definition}

It was shown in~\cite{Keppeler2020Answering} that, for SJFCQs that are not q-hierarchical, some complexity lower bounds apply to enumeration under updates and direct access under updates, even in the non-ranked versions.
These lower bounds are conditional to the \emph{OMv-conjecture}, which we introduce below.

\paragraph{OMv-conjecture} 
The \emph{Online Matrix-vector multiplication} (OMv) problem is defined as follows. We are given an $n \times n$ Boolean matrix $M$, on which we allow arbitrary polynomial preprocessing.
Then, a sequence of $n$ Boolean vectors $\vec{v}_1, \ldots, \vec{v}_n$ of dimension $n$ is given online, one vector at a time. For each $i \in  [n]$, we must output $M\vec{v}_i$ before receiving $\vec{v}_{i+1}$.
The complexity measure is the total running time required to output all vectors $M\vec{v}_i$. The \emph{OMv conjecture} states that no algorithm solves the OMv problem in time $O(n^{3-\varepsilon})$ for any $\varepsilon > 0$.

We now recall the lower bounds on enumeration under updates and direct access under updates from~\cite{Keppeler2020Answering}.
  
\begin{theorem}[{\cite[Theorem 3.13]{Keppeler2020Answering}}]\label{thm:UnrankedEnumlb}
Fix a number $\varepsilon > 0$ and a self-join-free conjunctive query $Q$. If $Q$ is not q-hierarchical, there is no algorithm with polynomial preprocessing time and $O(|D|^{1-\varepsilon})$ update time that enumerates $Q(D)$ with $O(|D|^{1-\varepsilon})$ delay in data complexity, unless the OMv-conjecture fails.
\end{theorem}

\begin{theorem}[{\cite[Theorem 3.16]{Keppeler2020Answering}}]\label{thm:UnrankedDAlb}
Fix a number $\varepsilon > 0$ and a self-join-free conjunctive query $Q$. If $Q$ is not q-hierarchical, there is no algorithm with polynomial preprocessing time and $O(|D|^{1-\varepsilon})$ update time that solves the direct access problem in time $O(|D|^{1-\varepsilon})$ delay in data complexity, unless the OMv-conjecture fails.
\end{theorem}

In contrast, it was also shown in~\cite{Keppeler2020Answering} that, for q-hierarchical SJFCQs, there are algorithms for efficient enumeration under updates and efficient direct access under updates.

\begin{theorem}[{\cite[Theorem 3.2]{Berkholz2017AnsweringCQ}}]
    There is an algorithm that receives a q-hierarchical SJFCQ $Q$ and a database instance $D$, and computes within $O(|Q|^{O(1)} |D|)$ preprocessing time a data structure that can be updated in time $O(|Q|^{O(1))}$ and allows the following:
    \begin{enumerate}
        \item enumerate $Q(D)$ with delay $O(|Q|^{O(1)})$
    \end{enumerate}
\end{theorem}

\begin{theorem}[{\cite[Theorem 3.9]{Keppeler2020Answering}}]
    There is an algorithm that receives a q-hierarchical SJFCQ $Q$ and a database instance $D$ and computes within preprocessing time $O(|Q|^{O(1)} |D| \log |D|)$ a data structure that can be updated in time $O(|Q|^{O(1)} \log |D|)$ and allows the following:
    \begin{enumerate}
        \item output the query result size $|Q(D)|$ in time $O(1)$ and,
        \item\label{item:enumUpper} enumerate the tuples in $Q(D)$ in lexicographic order according to the implementation order with delay $O(|Q|^{O(1)})$ and,
        \item upon input of an arbitrary $k$, takes time $O(|Q|^{O(1)} \log |D|)$ to immediately output the $k\ith$ answer that the enumeration procedure of (\ref{item:enumUpper}) would have output
    \end{enumerate}
\end{theorem}

In the next section, we will review proof techniques used to show these results because we will extend them for our own results.
In light of the lower bounds of \Cref{thm:UnrankedEnumlb} and \Cref{thm:UnrankedDAlb}, we will focus on q-hierarchical SJFCQs throughout the paper. Our goal is to investigate how this complexity picture is affected for ranked enumeration under updates and ranked direct access under updates as defined previously.

\section{Data Structures: Q-Trees, View-Trees, and Grounded-Trees} \label{sec:data_structure}

In this section, we 
define the data structure built in the preprocessing phase from the q-hierarchical SJFCQ and the database instance given as input. This data structure is from \cite{Kara2024TractableCQ}, and we call it a \emph{grounded-tree}.
As we study combined complexity, 
we present an algorithm
from \cite{Keppeler2020Answering} to build a so-called \emph{q-tree}
for a given q-hierarchical SJFCQ.
We then define from this the notions of \emph{view-trees} and \emph{grounded-trees}, and 
we show how to build grounded-trees from a q-hierarchical SJFCQ and a database instance. 

\subsection{From Q-trees to Grounded-Trees}

In this part, we define the data structure built in the preprocessing by giving the definitions of q-trees, view-trees, and grounded-trees.

We recall the definition of q-trees from \cite{Keppeler2020Answering}.

\begin{definition}
  \label{def:qtree}
    Let $Q$ be a q-hierarchical CQ. A \emph{q-tree} for $Q$ is a directed tree $T_Q = (V_Q,E_Q)$ with $V_Q = \vars(Q) \cup \{\vroot \}$ rooted in $\vroot$ where
    \begin{enumerate}
        \item for all atoms $\psi$ in $Q$, the set $\vars(\psi) \cup \{\vroot\}$ forms a directed path starting in $T_Q$ at $\vroot$, and
        \item $\free(Q) \cup \{\vroot \}$ is a connected subset in $T_Q$.
    \end{enumerate}
\end{definition}

A CQ $Q$ is q-hierarchical if and only if it admits a \mbox{q-tree} \cite[Lemma 4.2]{Keppeler2020Answering}. 
For each $v \in V_Q$, we denote by $\anc_{T_Q}(v)$ the ancestors of $v$ in $T_Q$, including $\vroot$. 
We can then define the following as in~\cite{Kara2024TractableCQ}:

\begin{definition}
    Let $Q$ be a q-hierarchical SJFCQ and $T_Q = (V_Q, E_Q)$ one of its q-trees.
    Let $T_Q^{\ext} = (V_Q^{\ext}, E_Q^{\ext})$ be the \emph{extended q-tree} of
    $T_Q$ which is defined as follows:
    we have $V_Q^{\ext} = V_Q \cup \{\psi \mid \psi \in Q\}$ and $E_Q^{\ext} = E_Q \cup \{(x_i, \psi_i) \mid x_i \in \vars(\psi_i) \text{ with } x_i \text{ the farthest variable from } \vroot \text{ in } T_Q \}$.  
    The \emph{view-tree} $V(T_Q)$ is the rooted tree with vertex set $\{V^{\Join}_{v} \mid v \in V_Q^{\ext}\} \cup\{V^{\Pi}_{(u,v)} \mid (u,v) \in E_Q^{\ext}\}$ and edge set $\{(V^{\Join}_{u}, V^{\Pi}_{(u,v)}),(V^{\Pi}_{(u,v)}, V^{\Join}_{v}) \mid (u,v) \in E_Q^{\ext},v \text{ a child of } u\}$, where each node 
    \(V^{\Join}_{v}\) or
    \(V^{\Pi}_{(u,v)}\) is associated to a view
    \(W^{\Join}_{v}\) or
    \(W^{\Pi}_{(u,v)}\) 
    which is a relational algebra expression defined as follows:
    \begin{enumerate}[label=(\roman*), ref=\roman*, leftmargin=*]
        \item for each leaf $v \in V_Q^{\ext}$ labeled by an atom $\psi \coloneq R(x_1, \ldots, x_{\ar(R)})$ then 
        the view associated to $V^{\Join}_{v}$ is
        $W^{\Join}_{v} \coloneq \psi$,
        \item for each internal vertex $v \in V_Q^{\ext}$ with children $v_1,\ldots,v_k$, then 
        the view associated to $V^{\Join}_{v}$ is
        $W^{\Join}_{v} \coloneq W^{\Pi}_{(v,v_1)} \Join \cdots \Join W^{\Pi}_{(v,v_k)}$,
        \item for each edge $(u,v) \in E_Q^{\ext}$, then 
        the view associated to $V^{\Pi}_{(u,v)}$ is
        $W^{\Pi}_{(u,v)} \coloneq \Pi_{(\anc_{T_Q}(u) \cup \{u\}) \setminus \{\vroot\}} W^{\Join}_{v}$.
    \end{enumerate}
\end{definition}

Given a database instance $D$, we now enrich the view-tree $V(T_Q)$ with the data of $D$, which gives a \emph{grounded-tree} $V_D(T_Q)$. Each vertex $V_v^{\Join}$ (resp.\ $V_{(u,v)}^{\Pi}$) in the view-tree $V(T_Q)$ is associated to a table $Z_v^{\Join}$ (resp.\ $Z_{(u,v)}^{\Pi}$) such that $Z_v^{\Join}$ (resp.\ $Z_{(u,v)}^{\Pi}$) is the evaluation of $W_v^{\Join}$ (resp.\ $W_{(u,v)}^{\Pi}$).
In particular, for every leaf $V_R^{\Join}$ labeled by an atom $R(x_1, \ldots, x_{\ar(R)})$, then $Z_R^{\Join} \coloneq R^D$. The other views of $V_D(T_Q)$ are then computed bottom-up from the leaf views, following their definitions.
Note that some projection views are defined on the same set of variables as their child. For instance, in~\Cref{fig:example_datastructure}, the view $W_{(x_2,R)}^{\Pi}$ is identical to $W_{R}^{\Join}$, therefore $Z_{(x_2,R)}^{\Pi} = Z_{R}^{\Join}$.
We do not try to remove these redundant views, because doing so would complicate the definitions and would not improve the asymptotic complexities of our algorithms.

In addition, whenever you have a vertex $V$ with associated view $W$ and a child $V_i$, then letting $Z$ and $Z_i$ be the tables respectively associated to them, for each tuple $t \in Z$, we denote by $\bucket{t}{Z_i}$ the set of tuples $t' \in Z_i$ such that $t'[\free(W)] = t$.
Note that for a variable $x_i$, the vertex $V_{x_i}^{\Join}$ has several children $V_{(x_i, x_j)}^{\Pi}$ in $V(T_Q)$. For each tuple $t \in Z_{x_i}^{\Join}$ and each $V_{(x_i, x_j)}^{\Pi}$ of $V_{x_i}$, we have that $\bucket{t}{Z_{(x_i, x_j)}^{\Pi}}$ has exactly one tuple. A projection view $V_{(x_i, x_j)}^{\Pi}$ has exactly one child view $V_{x_j}^{\Join}$. For each tuple $t \in Z_{(x_i, x_j)}^{\Pi}$, there exists $\bucket{t}{V_{x_j}^{\Join}}$ with several tuples $t'$ which are extensions of $t$ over $\anc_{T_Q}(x_j) \cup \{x_j\} \setminus{\{\vroot\}}$.

\begin{example}\label{example:db}

\Cref{fig:example_datastructure} represents an extended \mbox{q-tree} and its grounded-tree for the query $Q_1(x_1, x_2) = R(x_1,x_2) \land S(x_1, x_2) \land T(x_1)$, where by definition of a grounded-tree $Z_{R}^{\Join}(x_1,x_2) \coloneq R$, $Z_{S}^{\Join}(x_1, x_2) \coloneq S$ and $Z_{T}^{\Join}(x_1) \coloneq T$, and for the database instance $D$ given by the relations $R$, $S$ and $T$.

\begin{figure}[t]
    \centering
    \begin{subfigure}[c]{0.25\textwidth}
        \centering
        \begin{tikzpicture}[node/.style={minimum size=0.2cm, inner sep=2pt}]
            \node[node] (vstart) at (0,2) {$\vroot$};
            \node[node] (x1) at (0,1) {$x_1$};
            \node[node]  (x2) at (0,0) {$x_2$};
            \node[node, blue] (R) at (-1,-1) {$R(x_1, x_2)$};
            \node[node, blue] (S) at (1,-1) {$S(x_1, x_2)$};
            \node[node, blue] (T) at (1,0) {$T(x_1)$};

            \draw[->, thick] (vstart) -- (x1);
            \draw[->, thick] (x1) -- (x2);
            \draw[->, blue, thick] (x1) -- (T);
            \draw[->, blue, thick] (x2) -- (R);            
            \draw[->, blue, thick] (x2) -- (S);
        \end{tikzpicture}
        \subcaption{Extended q-tree $T_{Q_1}^{\ext}$}
    \end{subfigure}
    \hfill
    \begin{subfigure}[c]{0.4\textwidth}
        \centering
        \begin{tikzpicture}[node/.style={rectangle, thick, minimum size=0.2cm, inner sep=0pt, align=center, font=\footnotesize}]

            \node[node] (JoinVROOT) at (0,8) {\begin{tabular}{|c|}
                \hline
                $V_{\vroot}^{\Join}()$ \\
                \hline
                $()$ \\
                \hline
            \end{tabular}};
            \node[node] (ProjX1) at (0,7) {\begin{tabular}{|c|}
                \hline
                $V_{(\vroot,x_1)}^{\Pi}()$ \\
                \hline
                $()$ \\
                \hline
            \end{tabular}};
            \node[node] (JoinX1) at (0,5.8) {\begin{tabular}{|c|}
                \hline
                $V_{x_1}^{\Join}(x_1)$ \\
                \hline
                $b_1$ \\
                $b_3$ \\
                \hline
            \end{tabular}};
            \node[node] (ProjT) at (2,4.2) {\begin{tabular}{|c|}
                \hline
                $V_{(x_1,T)}^{\Pi}(x_1)$ \\
                \hline
                $b_1$ \\
                $b_3$ \\
                $b_4$ \\
                \hline
            \end{tabular}};
            \node[node] (JoinT) at (2,2.4) {\begin{tabular}{|c|}
                \hline
                $V_{T}^{\Join}(x_1)$ \\
                \hline
                $b_1$ \\
                $b_3$ \\
                $b_4$ \\
                \hline
            \end{tabular}};
            \node[node] (ProjX2) at (-2,4.2) {\begin{tabular}{|c|}
                \hline
                $V_{(x_1,x_2)}^{\Pi}(x_1)$ \\
                \hline
                $b_1$ \\
                \rowcolor{black!20}
                $b_3$ \\
                \hline
            \end{tabular}};
            \node[node] (JoinX2) at (-2,2.4) {\begin{tabular}{|cc|}
                \hline
                \multicolumn{2}{|c|}{$V_{x_2}^{\Join}(x_1, x_2)$} \\
                \hline
                $b_1$ & $c_1$ \\
                \rowcolor{black!10} 
                $b_3$ & $c_1$ \\
                \rowcolor{black!10}
                $b_3$ & $c_3$ \\
                \hline
            \end{tabular}};
            \node[node] (ProjR) at (-3.3,0) {\begin{tabular}{|cc|}
                \hline
                \multicolumn{2}{|c|}{$V_{(x_2,R)}^{\Pi}(x_1, x_2)$} \\
                \hline
                $b_1$ & $c_1$ \\
                $b_2$ & $c_2$ \\
                $b_3$ & $c_1$ \\
                $b_3$ & $c_3$ \\
                \hline
            \end{tabular}};
            \node[node] (JoinR) at (-3.3,-2) {\begin{tabular}{|cc|}
                \hline
                \multicolumn{2}{|c|}{$V_{R}^{\Join}(x_1, x_2)$} \\
                \hline
                $b_1$ & $c_1$ \\
                $b_2$ & $c_2$ \\
                $b_3$ & $c_1$ \\
                $b_3$ & $c_3$ \\
                \hline
            \end{tabular}};
            \node[node] (ProjS) at (-0.7,0) {\begin{tabular}{|cc|}
                \hline
                \multicolumn{2}{|c|}{$V_{(x_2,S)}^{\Pi}(x_1, x_2)$} \\
                \hline
                $b_1$ & $c_1$ \\
                $b_2$ & $c_1$ \\
                $b_3$ & $c_1$ \\
                $b_3$ & $c_3$ \\
                \hline
            \end{tabular}};
            \node[node] (JoinS) at (-0.7,-2) {\begin{tabular}{|cc|}
                \hline
                \multicolumn{2}{|c|}{$V_{S}^{\Join}(x_1, x_2)$} \\
                \hline
                $b_1$ & $c_1$ \\
                $b_2$ & $c_1$ \\
                $b_3$ & $c_1$ \\
                $b_3$ & $c_3$ \\
                \hline
            \end{tabular}};

            \draw[->, thick] (JoinVROOT.south) -- (ProjX1.north);
            \draw[->, thick] (ProjX1.south) -- (JoinX1.north);
            \draw[->, thick] (JoinX1.south) -- (ProjX2.north);
            \draw[->, thick] (ProjX2.south) -- (JoinX2.north);
            \draw[->, thick] (JoinX1.south) -- (ProjT.north);
            \draw[->, thick] (ProjT.south) -- (JoinT.north);
            \draw[->, thick] (JoinX2.south) -- (ProjR.north);
            \draw[->, thick] (ProjR.south) -- (JoinR.north);
            \draw[->, thick] (JoinX2.south) -- (ProjS.north);
            \draw[->, thick] (ProjS.south) -- (JoinS.north);
        \end{tikzpicture}
        \subcaption{Grounded-tree for $V(T_{Q_1})$}
    \end{subfigure}
    \hfill
    \begin{subfigure}[c]{0.25\textwidth}
        \centering
        \begin{tabular}{cc}
            \toprule
            $x_1$ & $x_2$ \\
            \midrule
            $b_1$ & $c_1$ \\
            $b_3$ & $c_1$ \\
            $b_3$ & $c_3$ \\
            \bottomrule
        \end{tabular}
        \subcaption{$Q_1(D)$}
    \end{subfigure}

    \caption{Representation of an extended q-tree, grounded-tree, and query result for $Q_1(D)$ (see \Cref{example:db})}
    \label{fig:example_datastructure}
\end{figure}

Note that the \mbox{q-tree} $T_Q$ is the subtree of \Cref{fig:example_datastructure} in black. In addition, for the tuple $t = (b_3)$ shown in gray, the tuples $t' \in \bucket{(b_3)}{V_{x_2}^{\Join}}$  are in a lighter gray. 
\end{example}

\subsection{Data Structure Implementation}
In this section, we first describe the implementation of the data structure defined previously, inspired by \cite{Keppeler2020Answering}, using lookup tables and $B^+$-trees. 
We then present the algorithms that we use when performing the preprocessing on this data structure.

\paragraph{$B^+$-trees.} Given a set of tuples $S$ and a ranking function $f$ that maps each tuple of $S$ to a value of an ordered domain, a \emph{B$^+$-tree for $(S,f)$} is a balanced search tree that stores $S$ in sorted order according to $f$. More precisely, all leaves are at the same depth, and each leaf stores a bounded number of consecutive tuples in the order $f$, whereas the internal nodes only store keys and pointers that guide the search for a specific tuple. Moreover, the leaves are chained in a linked list following $f$.
It is well known that $B^+$-trees admit algorithms allowing us to
insert a tuple, delete a tuple, retrieve a tuple, and retrieve the minimum tuple for $f$ each in time $O(\log |S|)$, since the height of the tree is logarithmic in the number of stored tuples. 
What is more, given a pointer to a tuple stored in the tree, accessing the tuple that follows it in the order induced by $f$ takes time $O(1)$, by moving to the next tuple of the same leaf or, at the end of a leaf, by following the pointer to the next leaf.

\paragraph{Grounded-tree implementation.} Let $V$ be a vertex of the view-tree $V(T_Q)$, let $W$ be the view associated to $V$, let $Z$ be the table associated to $V$, and let $V'$ be the parent of $V$. 
To navigate each table $Z$ efficiently, we group all tuples $t \in Z$ by their restriction $t'$ to $\free(W')$. We implement each table $Z$ as a lookup table whose keys are all tuples $t' \in Z'$ and whose value for each $t'$ is $\bucket{t'}{Z}$. 
For each $\bucket{t'}{Z}$, given a ranking function $f_{\bucket{t'}{Z}}$ that maps each $t \in \bucket{t'}{Z}$ to a value of an ordered domain, we store $\bucket{t'}{Z}$ in a $B^+$-tree sorted according to $f_{\bucket{t'}{Z}}$.

Therefore, given a table $Z$ and a key $t' \in Z'$, accessing the $B^+$-tree $\bucket{t'}{Z}$ takes constant time since $Z$ is a lookup table. Besides, given $\bucket{t'}{Z}$ and a ranking function $f_{\bucket{t'}{Z}}$, since each bucket is a $B^+$-tree for $(\bucket{t'}{Z},f_{\bucket{t'}{Z}})$, inserting or deleting a tuple $t$ in $\bucket{t'}{Z}$ while keeping it sorted according to $f_{\bucket{t'}{Z}}$ takes time $O(\log |D|)$. Further, given a pointer to a tuple in $\bucket{t'}{Z}$, retrieving its successor takes constant time.

\subsection{Preprocessing}

We next show how to build the grounded-tree from a SJFCQ $Q$. To do so, we first build a \mbox{q-tree} $T_Q$, extend it to a view-tree $V(T_Q)$, and, given a database instance $D$, we enrich $V(T_Q)$ into a grounded-tree with all the database facts from $D$.
We now recall the combined complexity of building a q-tree, given a q-hierarchical CQ.

\begin{lemma}[{\cite[Lemma 4.2]{Keppeler2020Answering}}]\label{lem:q-tree}
    A CQ $Q$ is q-hierarchical if and only if $Q$ has a q-tree. Moreover, there is an $O(|Q|^{O(1)})$ algorithm that decides whether an input CQ Q is q-hierarchical, and if so, outputs a q-tree.
\end{lemma}

Given a database instance $D$, a q-hierarchical SJFCQ $Q$, and a \mbox{q-tree} $T_Q$ for~$Q$, we now show how to build the grounded-tree for $T_Q$ in the preprocessing phase. We can show a preprocessing upper bound in $O(|D| \cdot |Q|)$ using the fact that we can sort in linear time in the RAM model \cite{Grandjean1996}.

\begin{proposition}\label{prop:linearPreproc}
    There exists an algorithm that, given a q-hierarchical SJFCQ $Q$, one of its q-trees $T_Q$, and a database instance $D$, builds the grounded-tree $V_D(T_Q)$, in which each bucket $b$ is sorted by a given ranking function $f_b$, in time $O(|D| |Q|)$.
\end{proposition}

Note that if one wants to avoid linear-time sorting algorithms, the
preprocessing can instead be performed in time $O(|D| \cdot |Q| \log |D|)$, using any
linearithmic sorting algorithm.

We can now prove \Cref{prop:linearPreproc}:

\begin{proof}[Proof of \Cref{prop:linearPreproc}]
    For a given q-hierarchical SJFCQ $Q$ and one of its q-trees $T_Q$, we first compute the extended \mbox{q-tree} $T_Q^{\ext}$ in time $O(|Q|)$ by adding, for each atom $\psi \in Q$, a leaf $\psi$ as a child of the farthest variable $x_i$ from $\vroot$ in $T_Q$ such that $x_i \in \vars(\psi)$. 
    Afterwards, we compute the view-tree $V(T_Q)$ by traversing $T_Q^{\ext}$ while adding $V_{v}^{\Join}$ for each vertex $v \in T_Q^{\ext}$, and $V_{(u, v)}^{\Pi}$ for each edge $(u, v) \in T_Q^{\ext}$, in time $O(|Q|)$. We then initialize all tables to the empty table by creating an empty table $Z$ for each vertex $V$ of the view-tree, giving us the empty grounded tree $V_D(T_Q)$.
    Given a database instance $D$, we then set $Z_R^{\Join} \coloneq R^D$ for every relation $R$ of $D$. We then propagate the tuples bottom-up as follows. Let $V$ be a vertex, let $V_i$ be a child of $V$, let $Z$ and $Z_i$ be the tables associated to $V$ and $V_i$, and let $W$ be the view associated to $V$. For each table $Z_i$, we evaluate the view $W_i$ and group the tuples of $Z_i$ into buckets according to their restriction to $\free(W)$: each tuple $t' \in Z_i$ is placed in the bucket $\bucket{t}{Z_i}$ with key $t = t'[\free(W)]$.
    If $W$ is a projection view, then $Z$ contains the keys $t$ of the buckets of its only child $Z_i$. If $W$ is a join view, we fix one child $V_i$ and scan the keys $t$ of the buckets of $Z_i$; for each of them, we check in constant time per other child $V_j$ that $\bucket{t}{Z_j}$ exists, and $t$ is added to $Z$ only if all these buckets exist.
    Every table has at most $|D|$ tuples: the leaves are relations of $D$, a projection view has at most as many tuples as its child, and a join view has at most as many tuples as any of its children. Hence, a projection vertex is processed in time $O(|D|)$, and a join vertex in time $O(|D|)$ per child. Since the sum of the number of children over all vertices is the number of edges of the view-tree, which is $O(|Q|)$, all views are evaluated in time $O(|Q| \cdot |D|)$.
    Using the fact that we can sort in linear time in the RAM model \cite{Grandjean1996}, we then sort each $\bucket{t}{Z_i}$ according to its ranking function $f_{\bucket{t}{Z_i}}$ and build, for each bucket, a $B^+$-tree from the sorted list in linear time. Each table $Z$ contains at most $|D|$ tuples in total, separated into several buckets, so this takes $O(|D|)$ per table, and $O(|Q| \cdot  |D|)$ overall.
    Hence, there exists an algorithm that builds a grounded tree $V_D(T_Q)$, in which each bucket $\bucket{t'}{Z}$ is sorted according to $f_{\bucket{t'}{Z}}$, in time $O(|Q| \cdot |D|)$.
\end{proof}

For the query evaluation problems we tackle throughout this paper, we can therefore build a grounded-tree for a q-hierarchical SJFCQ $Q$ in time $O(|Q|^{O(1)} + |Q| \cdot |D|)$, i.e., linear preprocessing in data complexity, by building one of its q-trees with \Cref{lem:q-tree} and applying \Cref{prop:linearPreproc}. 

\section{Hierarchy-Consistent Orders}\label{sec:HC}

We now define a notion of order on the free variables of a q-hierarchical CQ.

\begin{definition}
    \label{def:query-topo}
    Let $Q$ be a q-hierarchical CQ. Then, $Q$ induces a partial order on $\free(Q)$ where $x_i <_Q x_j$ if $\atoms_Q(x_j) \subsetneq \atoms_Q(x_i)$. A total order $\prec$ on $\free(Q)$ is \emph{hierarchy-consistent} for $Q$ if it is a linear extension of this partial order.
\end{definition}

Given a q-hierarchical CQ $Q$, and a hierarchy-consistent order $\prec$ over $\free(Q)$, we now want to perform ranked enumeration and ranked direct access. 
To do so, we need to build a data structure adapted to $\prec$. We therefore want a \mbox{q-tree}, used to build the data structure, in which $\prec$ can be obtained easily, for instance by a topological sort. It follows from the proof of \cite[Lemma 4.2]{Keppeler2020Answering} that several q-trees may
represent a q-hierarchical CQ $Q$. However, every hierarchy-consistent order $\prec$ cannot be obtained by a topological sort of any \mbox{q-tree}. 
Indeed, for each pair $x_1, x_2 \in \vars(Q)$,
whenever $\atoms_Q(x_1) = \atoms_Q(x_2)$ and $x_1 \in \free(Q) \iff x_2 \in \free(Q)$, the ancestor-descendant relation between these variables may vary: in one \mbox{q-tree} $x_2$ is an ancestor of $x_1$, while in another $x_1$ is an ancestor of $x_2$, as shown in \Cref{fig:diff_q-trees}. Thus, a
hierarchy-consistent order where $x_1 < x_2$ cannot be obtained by a topological
sort of the first q-tree. We make this clearer with an example:

\begin{example}
    Let $Q(x_1, x_2, x_3, x_4) \coloneq R(x_1, x_2, x_4) \land S(x_3, x_4)$ be a q-hierarchical CQ. The linear extension $\prec$ such that $\vroot \prec x_4 \prec x_3 \prec x_2 \prec x_1$ is a topological sort of $T_1$ but cannot be obtained by a topological sort of $T_2$ in \Cref{fig:diff_q-trees}. 

    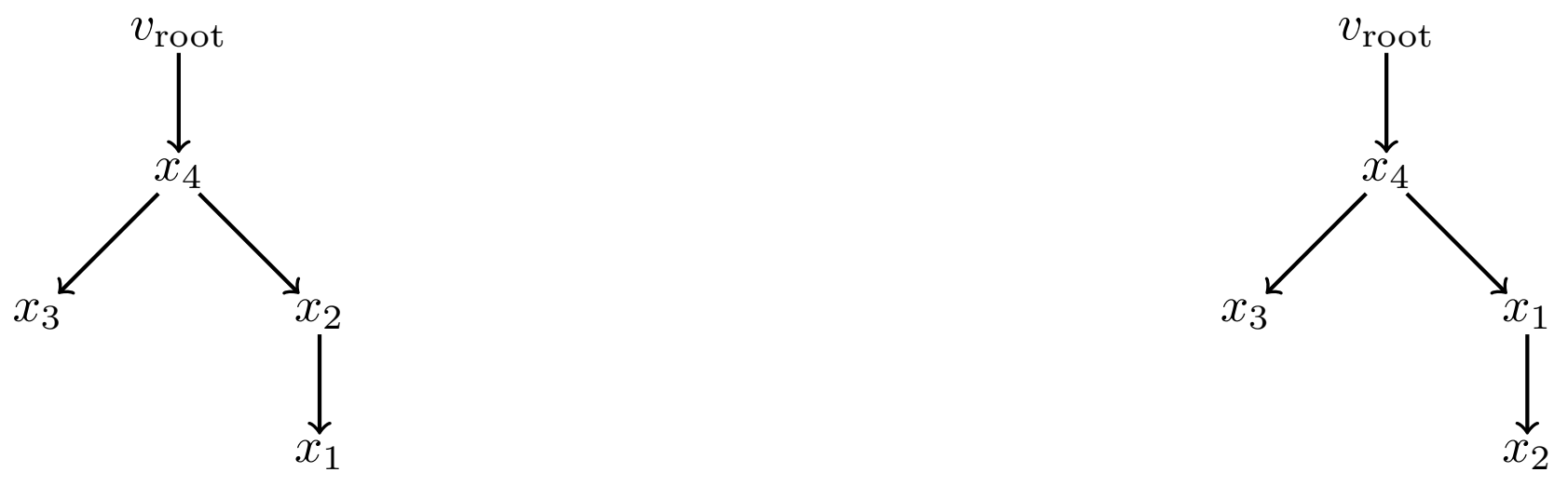
\begin{figure}[htbp]
\centering

\begin{subfigure}[t]{0.45\textwidth}
\centering
\begin{tikzpicture}[node/.style={minimum size=0.2cm, inner sep=1pt}]
	\node[node] (vstart) at (0,2) {$\vroot$};
	\node[node] (x4) at (0,1) {$x_4$};
	\node[node] (x3) at (-1,0) {$x_3$};
	\node[node] (x2) at (1,0) {$x_2$};
    \node[node] (x1) at (1,-1) {$x_1$};
	
	\draw[->, thick] (vstart) -- (x4);
    \draw[->, thick] (x4) -- (x3);
	\draw[->, thick] (x4) -- (x2);
	\draw[->, thick] (x2) -- (x1);
\end{tikzpicture}
\subcaption{$T_1$}
\end{subfigure}
\hfill
\begin{subfigure}[t]{0.45\textwidth}
\centering
\begin{tikzpicture}[node/.style={minimum size=0.2cm, inner sep=1pt}]
	\node[node] (vstart) at (0,2) {$\vroot$};
    \node[node] (x4) at (0,1) {$x_4$};
	\node[node] (x3) at (-1,0) {$x_3$};
	\node[node] (x1) at (1,0) {$x_1$};
    \node[node] (x2) at (1,-1) {$x_2$};

    \draw[->, thick] (vstart) -- (x4);
	\draw[->, thick] (x4) -- (x3);
	\draw[->, thick] (x4) -- (x1);
	\draw[->, thick] (x1) -- (x2);
\end{tikzpicture}
\subcaption{$T_2$}
\end{subfigure}
\caption{Q-trees for $Q(x_1, x_2, x_3, x_4) \coloneq R(x_1, x_2, x_4) \land S(x_3, x_4)$}
\label{fig:diff_q-trees}
\end{figure}

\end{example}

Hence, in the preprocessing phase, we build a \emph{compatible} \mbox{q-tree} for the order $\prec$, defined as follows.

\begin{definition}
    Let $Q$ be a q-hierarchical CQ, $T_Q$ one of its q-trees, and $\prec$ a hierarchy-consistent order on $\free(Q)$. We say $T_Q$ is \emph{compatible} with $\prec$ if $\prec$ is the restriction to $\free(Q)$ of a topological order of $T_Q$.
\end{definition}

We now show that such a \mbox{q-tree} exists, strengthening \cite[Lemma 4.2]{Keppeler2020Answering} (which we had restated previously as \Cref{lem:q-tree}).

\begin{claim}\label{claim:compatible-qtree}
    Given a q-hierarchical CQ $Q$ and a hierarchy-consistent order $\prec$ on its free variables, we can build a \mbox{q-tree} $T_Q$ compatible with $\prec$ in $O(|Q|^{O(1)})$.
\end{claim}

\begin{proof}
    To build such a q-tree compatible with a hierarchy-consistent order $\prec$, we adapt the algorithm from \cite[Lemma 4.2]{Keppeler2020Answering} using $\prec$ to break ties when, for $x_1, x_2 \in \free(Q)$, we have $\atoms_Q(x_1) = \atoms_Q(x_2)$. This algorithm builds a \mbox{q-tree} $T_Q$ of $Q$ such that each path of $T_Q$ is a path of the form $\vroot, x_1, \ldots, x_s$ where $\atoms_Q(x_s) \subseteq \cdots \subseteq \atoms_Q(x_1)$. In this algorithm, when for two variables $x_1, x_2 \in \free(Q)$, we have $\atoms_Q(x_1) = \atoms_Q(x_2)$, we choose arbitrarily which variable will be the ancestor of the other. To get a \mbox{q-tree} in which $\prec$ is a topological sort, we break this tie following the order $\prec$, i.e., when for two variables $x_1, x_2 \in \free(Q)$, we have $\atoms_Q(x_1) = \atoms_Q(x_2)$, if $x_1 \prec x_2$ then $x_1$ is an ancestor of $x_2$, otherwise $x_2$ is an ancestor of $x_1$.
\end{proof}

Before showing how to achieve ranked enumeration (resp.\ direct access) under updates for a hierarchy-consistent given order, we first show that, when the database changes due to the insertion or deletion of a tuple, we can update the grounded-tree in time $O(\log |D|)$.
Recall that in a grounded-tree, each bucket $b$ is sorted by a ranking function $f_b$. When the order $\prec$ used to build the grounded-tree is hierarchy-consistent, we define these ranking functions as the \emph{lexicographic ranking functions} induced by $\prec$.

Let $V$ be a non-root vertex of the grounded tree $V_D(T_Q)$, let $W$ be the view associated to $V$, let $Z$ be the table associated to $V$, and let $V'$ be the parent of $V$ with associated view $W'$.
Let $Q$ be a q-hierarchical CQ, 
let $\prec$ be the hierarchy-consistent order over $\free(Q)$, 
let $T_Q$ be a q-tree for $Q$ compatible with $\prec$, and let
$D$ be a database instance. We extend $\prec$ arbitrarily to a topological order of $T_Q$ over all of $\vars(Q)$.
For each bucket $\bucket{t'}{Z}$ of the grounded tree $V_D(T_Q)$, the lexicographic ranking
function $f_{\bucket{t'}{Z}}$, induced by $\prec$ over $\vars(Q)$, is simply the identity: $f_{\bucket{t'}{Z}}(t) \coloneq t$ for every $t \in \bucket{t'}{Z}$, where the domain is the set of tuples over $\free(W)$ ordered lexicographically according to $\prec$.
In other words, for all $t_1, t_2 \in \bucket{t'}{Z}$, we have $f_{\bucket{t'}{Z}}(t_1) \precbar f_{\bucket{t'}{Z}}(t_2)$ if and only if $t_1 \precbar t_2$, where $\precbar$ is the lexicographic order induced by $\prec$.

\begin{lemma}\label{lem:logUpdate-HC}
    Let $Q$ be a q-hierarchical SJFCQ, $\prec$ a hierarchy-consistent order over $\free(Q)$, $T_Q$ a \mbox{q-tree} for $Q$ compatible with $\prec$, and $D$ a database instance. Let $V_D(T_Q)$ be the grounded-tree for $T_Q$ and $D$ whose buckets $b$ are sorted using the lexicographic ranking functions $f_b$ defined by $\prec$. Given an update command $\texttt{update}~R(a_1,\dots, a_{\ar(R)})$, let $D'$ be the database instance obtained by applying $\texttt{update}~R(a_1,\dots, a_{\ar(R)})$ to $D$. 
    There exists an algorithm that, given $V_D(T_Q)$ and given an $\texttt{update}~R(a_1,\dots, a_{\ar(R)})$, computes in time $O(|Q| \log |D|)$ the grounded tree $V_{D'}(T_Q)$, whose buckets $b$ are sorted using the lexicographic ranking functions $f_b$.
\end{lemma}

\begin{proof}
    Let $V_D(T_Q)$ be a grounded-tree and for each bucket of $V_D(T_Q)$, let its tuples be sorted lexicographically. 
    Let $\texttt{update}~R(a_1,\dots,a_{\ar(R)})$ be an update of a tuple $t \coloneq R(a_1, \ldots, a_{\ar(R)})$, and let $D'$ be the database instance obtained by applying $\texttt{update}~R(a_1,\dots, a_{\ar(R)})$ to $D$. We build $V_{D'}(T_Q)$ by inserting or removing database facts from $V_D(T_Q)$ as follows. We distinguish two cases depending on the kind of update.

    When the update is an insertion, for the table $Z_{R}^{\Join}(X \cup Y)$
    with $X$ the set of free variables of the view $W_{(x_i, R)}^{\Pi}$, we insert $t$ in the bucket $\bucket{t[X]}{Z_{R}^{\Join}}$. Accessing $\bucket{t[X]}{Z_{R}^{\Join}}$ takes constant time, and inserting $t$ in $\bucket{t[X]}{Z_{R}^{\Join}}$ while keeping $\bucket{t[X]}{Z_{R}^{\Join}}$ sorted lexicographically takes logarithmic time. Then, we propagate the insertion upwards in $V_D(T_Q)$ on the path from $V_{R}^{\Join}$ to $V_{\vroot}^{\Join}$ as follows. 
    When we insert a tuple $t$ in a $\bucket{t'}{Z_{x_j}^{\Join}}$, with $t' = t[W_{(x_i, x_j)}^{\Pi}]$, we insert $t'$ in $\bucket{t''}{Z_{(x_i, x_j)}^{\Pi}}$ where $t'' = t'[\free(W_{(x_i, x_j)}^{\Pi})]$. If $t'$ already exists in $\bucket{t''}{Z_{(x_i, x_j)}^{\Pi}}$, the propagation stops. 
    Otherwise, when we insert a tuple $t$ in $\bucket{t'}{Z_{(x_i, x_j)}^{\Pi}}$ with $t' = t[W_{x_i}^{\Join}]$, we insert $t'$ in $\bucket{t''}{Z_{x_i}^{\Join}}$ where $t'' = t'[W_{x_i}^{\Join}]$, if and only if $t$ exists in all $Z_{(x_i, x_k)}^{\Pi}$ for each child $x_k$ of $x_i$ in the extended q-tree. It is equivalent to checking if $\bucket{t}{Z_{(x_i, x_k)}^{\Pi}}$ exists, which takes constant time. 
    In the worst case, we propagate an update until $V_{\vroot}^{\Join}$, which takes $O(|Q| \log |D|)$.

    When the update is a deletion, for the table $Z_{R}^{\Join}(X \cup Y)$ with $X$ the set of free variables of the view $W_{(x_i, R)}^{\Pi}$, we remove $t$ from $\bucket{t[X]}{Z_{R}^{\Join}}$. Accessing $\bucket{t[X]}{Z_{R}^{\Join}}$ takes constant time, and removing $t$ from $\bucket{t[X]}{Z_{R}^{\Join}}$ while keeping $\bucket{t[X]}{Z_{R}^{\Join}}$ sorted lexicographically takes logarithmic time.
    Let $t \coloneq (a_1, \ldots, a_{\ar(R)})$ be the deleted tuple. Then, we propagate the deletion upwards in $V_D(T_Q)$ on the path from $V_{R}^{\Join}$ to $V_{\vroot}^{\Join}$ as follows.
    When we delete a tuple $t$ from a $\bucket{t'}{Z_{x_j}^{\Join}}$, with $t' = t[\free(W_{(x_i, x_j)}^{\Pi})]$, if $\bucket{t'}{Z_{x_j}^{\Join}}$ becomes empty, we remove $t'$ from $\bucket{t''}{Z_{(x_i, x_j)}^{\Pi}}$, which takes time $O(\log |D|)$. If there still exists a tuple in $\bucket{t'}{V_{x_j}^{\Join}}$ the propagation stops. When we delete a tuple $t$ in $\bucket{t'}{Z_{(x_i, x_j)}^{\Pi}}$, with $t' = t[\free(Z_{x_i}^{\Join})]$, we remove $t'$ from $\bucket{t''}{Z_{x_i}^{\Join}}$, where $t'' = t'[\anc_{T_Q}(x_i)]$ if it exists; otherwise, the propagation stops. 
    In the worst case, we propagate a deletion until $V_{\vroot}^{\Join}$, which takes $O(|Q| \log |D|)$.
    Hence, an update takes $O(|Q| \log |D|)$.
\end{proof}

We can now show how to achieve ranked enumeration under updates when the order in the input is hierarchy-consistent.

\subsection{Ranked Enumeration Under Updates}

In this part, given a q-hierarchical SJFCQ $Q$ and a database instance $D$, we propose an enumeration algorithm in $\langle |D| |Q|, |Q|, |Q| \log |D| \rangle$ delay in combined complexity, i.e., linear preprocessing, logarithmic update, and constant delay times in data complexity, when the order on the free variables of $Q$ is hierarchy-consistent.

\begin{theorem}
    There exists an algorithm that, given a q-hierarchical SJFCQ $Q$, a hierarchy-consistent total order $\prec$ on its free variables, a q-tree $T_Q$ for $Q$ compatible with $\prec$ and a database instance $D$, enumerates $Q(D)$ in the lexicographic order defined by $\prec$ with $O(|D| |Q|)$ preprocessing, $O(|Q| \log |D|)$ update time, and $O(|Q|)$ delay.
\end{theorem}

\begin{proof}
Given a q-hierarchical SJFCQ $Q$, a database instance $D$, and a hierarchy-consistent order $\prec$ over $\free(Q)$, we apply the preprocessing procedure to build the grounded-tree $V_D(T_Q)$ using \Cref{prop:linearPreproc}. Then, when updates arrive, we handle them using \Cref{lem:logUpdate-HC}. Hence, all that remains is to explain how to perform ranked enumeration, which we now present.

Let $x_1, \ldots, x_l$ be the free variables of $Q$ listed according to $\prec$. 
\Cref{alg:RankedEnum} is written as nested loops over $x_1,\ldots,x_l$ for readability. 
In this algorithm, we traverse the grounded-tree $V_D(T_Q)$ from top to bottom to build each answer $t_{Q(D)}$ of the enumeration.
For a non-root variable $x_j$, the bucket $\bucket{t'}{Z_{x_j}^{\Join}}$ to traverse is determined by a tuple $t'$ chosen in table $Z_{(x_i,x_j)}^{\Pi}$. Recall that there is only one tuple $t'$ per bucket $\bucket{t}{Z_{(x_i, x_j)}^{\Pi}}$ such that $t \in Z_{x_i}^{\Join}$ and $t = t'$, therefore $\bucket{t'}{Z_{x_j}^{\Join}} = \bucket{t}{Z_{x_j}^{\Join}}$. 
In the following, we traverse the grounded-tree by only visiting the join vertices $V_{x_i}^{\Join}$.
Moreover, since $\prec$ is hierarchy-consistent, $t$ is chosen before the tuple in $\bucket{t'}{Z_{x_j}^{\Join}}$, so $\bucket{t'}{Z_{x_j}^{\Join}}$ is always known when we need it.

To enumerate the first answer of $Q(D)$, we start from the root table $Z_{\vroot}^{\Join}$. Let $t_0 \coloneq ()$ be the tuple of $Z_{\vroot}^{\Join}$, if it exists. Otherwise, we return \texttt{end-of-enumeration}. Then, for every child $c$ of $\vroot$ in $T_Q$, we save the bucket $\bucket{t_0}{Z_{c}^{\Join}}$ in a variable $\texttt{bucket}[c]$.
Afterward, for $j = 1, \ldots, l$ in this order, we take the first tuple $t_j$ of the bucket saved in variable $\texttt{bucket}[x_j]$, determined by the tuple chosen for its parent. At last, we output the answer built from the first tuple $t_l$ of the corresponding bucket, and pause the algorithm with the command \texttt{yield}.

To enumerate the next answer of $Q(D)$, we look for the rightmost variable $x_i$ whose current tuple is not the last of its bucket. If there is none, the enumeration ends. Otherwise, we replace the tuple of $x_i$ with the following tuple of its bucket, and for each $x_j$ with $j > i$, in the order $\prec$, we take the first tuple of the bucket determined by the tuple chosen for its parent. Only the variables after $x_{i-1}$ change between two consecutive answers.

Getting $\bucket{t}{Z_{x_j}^{\Join}}$ for a tuple $t$ and a variable $x_j$ takes constant time, since each vertex $V_{x_j}^{\Join}$ is a look-up table where keys are all assignments of the free variables of $W_{x_i}^{\Join}$ where $x_i$ is the parent of $x_j$ in the extended \mbox{q-tree} $T_Q^{\ext}$. Therefore, the delay to output a new tuple is at most the traversal of every join vertex of the grounded-tree until the last free variable of the order $\prec$, i.e., the delay of \Cref{alg:RankedEnum} is $O(|Q|)$.
Combining the delay bound with the bounds from \Cref{prop:linearPreproc} and \Cref{lem:logUpdate-HC}, there exists an algorithm that enumerates $Q(D)$ with $O(|D| \cdot |Q|)$ preprocessing, $O(|Q| \log |D|)$ update time and $O(|Q|)$ delay.
\end{proof}

\begin{algorithm}
    \DontPrintSemicolon
    \SetKw{Yield}{yield}
    \SetKwProg{Fn}{Fn}{:}{}
    \KwIn{$V_D(T_Q)$: a grounded-tree for $Q(D)$}
    \KwOut{answers of $Q(D)$ in the lexicographic order induced by $\prec$}

    \Begin{
        Let $x_1 < \ldots < x_l$ be the variables of $\free(Q)$ in the desired hierarchy-consistent order $\prec$\;
        \For{$t_0 \in Z_{\vroot}^{\Join}$}{
            $t_{Q(D)} \gets ()$\;
            \For{$c$ child of $\vroot$ in $\free(Q)$}{
                $\texttt{bucket}[Z_{c}^{\Join}] \gets \bucket{t_0}{Z_{c}^{\Join}}$
            }
            \For{$t_1 \in \texttt{bucket}[Z_{x_1}^{\Join}]$}{
                $t_{Q(D)}[x_1] \gets t_1[x_1]$\;
                \For{$c$ child of $x_1$ in $\free(Q)$}{
                    $\texttt{bucket}[Z_{c}^{\Join}] \gets \bucket{t_1}{Z_{c}^{\Join}}$
                }
                \vdots\;
                \For{$t_l \in \texttt{bucket}[Z_{x_l}^{\Join}]$}{
                    $t_{Q(D)}[x_l] \gets t_l[x_l]$\;
                    \Yield $t_{Q(D)}$\;
                }
            }
        }
        \Yield \texttt{end-of-enumeration}\;
    }
    \caption{Ranked enumeration according to a hierarchy-consistent order $\prec$}\label{alg:RankedEnum}
\end{algorithm}

\subsection{Ranked Direct Access Under Updates}

In this part, we first explain how to enrich the grounded-tree representing a database instance for a given hierarchy-consistent order on the free variables of a q-hierarchical SJFCQ $Q$. Then, we adapt an algorithm from \cite{Carmeli2023} into an algorithm in $\langle |D| |Q|, |Q| \log |D|, |Q| \log |D| \rangle$ that, given $Q$, a hierarchy-consistent order $\prec$, a \mbox{q-tree} for $T_Q$ compatible with $\prec$, and a database instance, answers ranked direct access under updates in $O(|Q| \log(|D|))$ access time.

\paragraph{Enriched grounded-tree for ranked direct access under updates}
Let $V_D(T_Q)$ be a grounded-tree as described in \Cref{sec:data_structure} for a given q-hierarchical SJFCQ $Q$, a given database instance $D$, and a hierarchy-consistent order $\prec$. Recall that each bucket $b$ is sorted according to a lexicographic ranking function $f_b$. This ranking function is the identity: $f_b(t) \coloneq t$ for every $t \in b$, meaning $b$ is sorted lexicographically.
We call $V_D^{\free}(T_Q)$ the subtree of $V_D(T_Q)$ containing all vertices $\{ V_{x_i}^{\Join} \cup V_{(x_i, x_j)}^{\Pi} \mid x_i, x_j \in \free(Q) \cup \{\vroot\}\}$.
As $Q$ is q-hierarchical, $V_D^{\free}(T_Q)$ is a connected subtree.
Let $V$ be a vertex, let $V_i$ be the parent of $V$, let $Z$ and $Z_i$ be the tables associated to $V$ and $V_i$, and let $W$ be the view associated to $V$.
For each tuple $t$ in a $\bucket{t_i}{Z}$ with $t_i \in Z_i$, we add weights $w_{\bucket{t_i}{Z}}(t)$.
Note that these weights are different from the ranking function $f_b$, as they are not used to sort the tuples in a bucket. The weight $w_{\bucket{t_i}{Z}}(t)$ represents the number of different answers in $Q(D)$ that extend $t$ over the variables in the views of the subtree rooted in $V$ in $V_D^{\free}(T_Q)$.
We compute these weights bottom-up while building the grounded-tree $V_D(T_Q)$, and when an update occurs. For leaves in the subtree $V_D^{\free}(T_Q)$, we initiate all weights to $1$. When we propagate the tuple upward in $V_D(T_Q)$, as explained in \Cref{prop:linearPreproc} and \Cref{lem:logUpdate-HC}, we compute the weight for each tuple propagated as follows.
Note that in this case, the propagation of an update never stops before achieving $V_{\vroot}^{\Join}$, since we need to update weights for the parent of each vertex. This still takes $O(|Q| \log |D|)$.

Let $x_i, x_j \in \free(Q) \cup \{\vroot\}$ such that $x_j$ is a child of $x_i$ in $T_Q$, and $t_i$ the tuple restricted over the variable of the view $W_i$ associated to the parent $V_i$ of vertex $V$.
\[
w_{\bucket{t_i}{Z}}(t) \coloneq
\begin{cases}
    1 & \text{if  } Z \text{ is associated to a leaf in } V_D^{\free}(T_Q)\\
    \sum\limits_{t'}w_{\bucket{t}{Z_{x_j}^{\Join}}}(t') & \text{if } Z = Z_{(x_i, x_j)}^{\Pi}, \\
    \prod\limits_{x_j} w_{\bucket{t}{Z_{(x_i, x_j)}^{\Pi}}}(t') & \text{if } Z = Z_{x_i}^{\Join}.
\end{cases}
\]

Therefore, if $t \in Z_{(x_i, x_j)}^{\Pi}$, its weight $w_b(t)$ is the sum of all the weights of the tuples $t'$ that are extensions of $t$ in the table $Z_{x_j}^{\Join}$. Otherwise, recall that $\bucket{t}{Z_{(x_i, x_j)}^{\Pi}}$ contains exactly one tuple. Then, if $t \in Z_{x_i}^{\Join}$, its weight $w_b(t)$ is the product, over all children $x_j$ of $x_i$, of the weight of the unique tuple $t_{x_j} \in \bucket{t}{Z_{(x_i, x_j)}^{\Pi}}$.

In addition to $w_b(t)$, we compute a prefix sum $\texttt{start}_b(t)$ for each bucket $b$ and each tuple $t \in b$.
This sum is the cumulative sum of the weights of the tuples that strictly precede $t$, i.e., $\texttt{start}_b(t) \coloneq \sum\limits_{t' \in b, t' < t} w(t')$.

\begin{example}
  \label{exa:q2}
    We now consider the query $Q_2(x_1, x_3) \coloneq R(x_1, x_2) \land T(x_3)$, where by definition of a view-tree we have $Z_{R}^{\Join} \coloneq R^D$ and $Z_{T}^{\Join} \coloneq T^D$.

    We enrich the view-tree $V(T_{Q_2})$ with the weights $w(t)$ and the prefix sum $\texttt{start}(t)$. 
    The result is represented in \Cref{fig:example_datastructure_weigth}.
    
    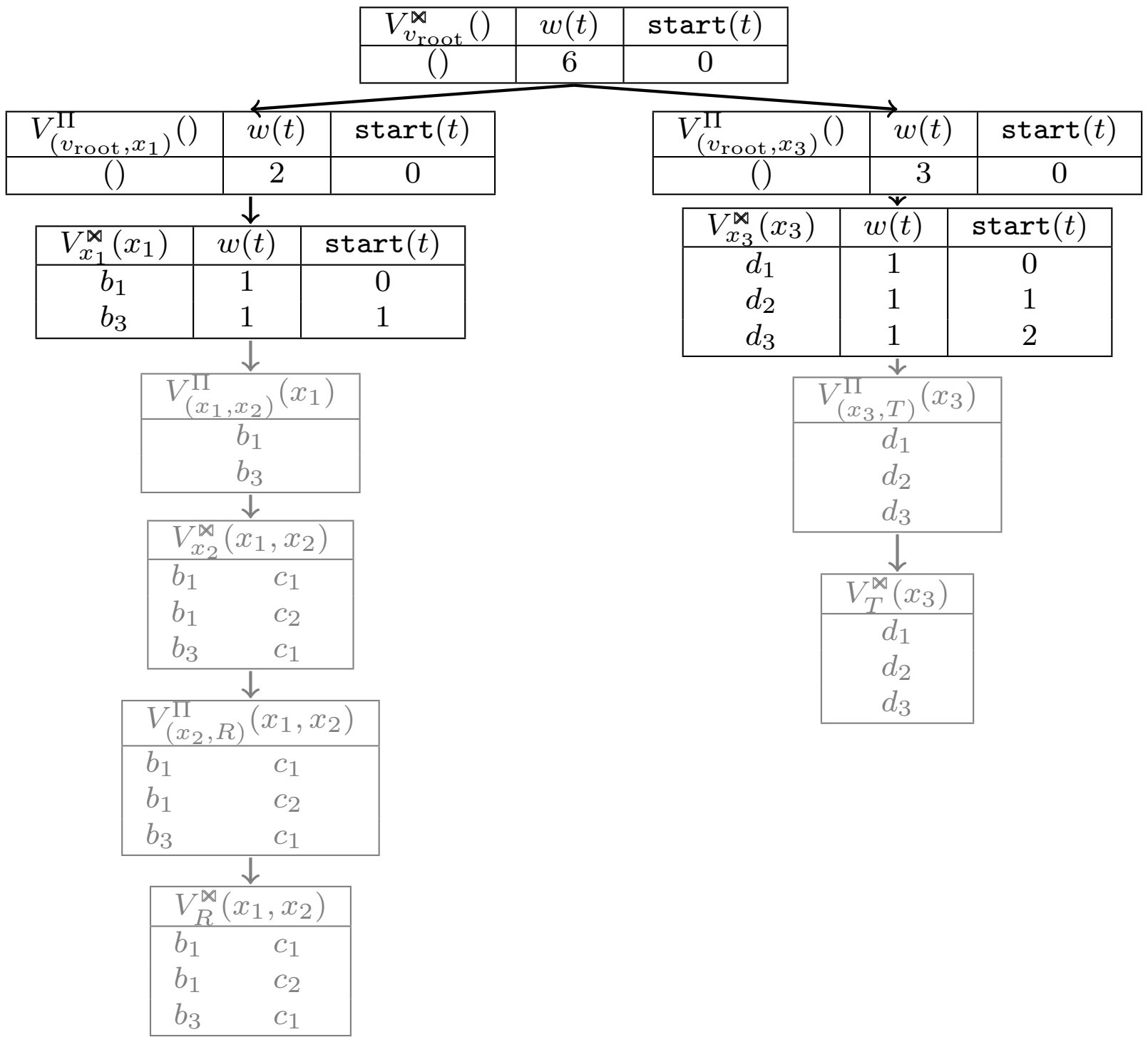
\begin{figure}[ht]
    \centering
    \begin{minipage}[c]{0.25\textwidth}
        \centering
        \begin{subfigure}[c]{\textwidth}
            \centering
            \begin{tikzpicture}[node/.style={minimum size=0.2cm, inner sep=2pt}]
                \node[node] (vstart) at (0,2) {$\vroot$};
                \node[node] (x1) at (-1,1) {$x_1$};
                \node[node] (x2) at (-1,0) {$x_2$};
                \node[node] (x3) at (1,1) {$x_3$};

                \node[node, blue] (R) at (-1,-1) {$R(x_1, x_2)$};
                \node[node, blue] (T) at (1,0) {$T(x_3)$};
     
                \draw[->, thick] (vstart) -- (x1);
                \draw[->, thick] (x1) -- (x2);
                \draw[->, thick] (vstart) -- (x3);
                \draw[->, blue, thick] (x3) -- (T);
                \draw[->, blue, thick] (x2) -- (R);            
            \end{tikzpicture}
            \subcaption{Extended q-tree $T_{Q_2}$}
        \end{subfigure}
    
        \vspace{1cm}
    
        \begin{subfigure}[t]{\textwidth}
            \centering
            \begin{tabular}{cc}
                \toprule
                $x_1$ & $x_3$ \\
                \midrule
                $b_1$ & $d_1$ \\
                $b_1$ & $d_2$ \\
                $b_1$ & $d_3$ \\
                $b_3$ & $d_1$ \\
                $b_3$ & $d_2$ \\
                $b_3$ & $d_3$ \\
                \bottomrule
            \end{tabular}
            \subcaption{$Q_2(D)$}
        \end{subfigure}
    \end{minipage}
    \hfill
    \begin{subfigure}[c]{0.72\textwidth}
        \centering
        \begin{tikzpicture}[node/.style={rectangle, thick, minimum size=0.2cm, inner sep=0pt, align=center, font=\footnotesize}]

            \node[node] (JoinVROOT) at (0,8) {\begin{tabular}{|c|c|c|}
                \hline
                $V_{\vroot}^{\Join}()$ & $w(t)$ & $\texttt{start}(t)$ \\
                \hline
                $()$ & $6$ & $0$ \\
                \hline
            \end{tabular}};
            \node[node] (ProjX1) at (-3,7) {\begin{tabular}{|c|c|c|}
                \hline
                $V_{(\vroot,x_1)}^{\Pi}()$ & $w(t)$ & $\texttt{start}(t)$ \\
                \hline
                $()$ & $2$ & $0$ \\
                \hline
            \end{tabular}};
            \node[node] (JoinX1) at (-3,5.8) {\begin{tabular}{|c|c|c|}
                \hline
                $V_{x_1}^{\Join}(x_1)$ & $w(t)$ & $\texttt{start}(t)$ \\
                \hline
                $b_1$ & $1$ & $0$ \\
                $b_3$ & $1$ & $1$ \\
                \hline
            \end{tabular}};
            \node[node] (ProjX3) at (3,7) {\begin{tabular}{|c|c|c|}
                \hline
                $V_{(\vroot,x_3)}^{\Pi}()$ & $w(t)$ & $\texttt{start}(t)$ \\
                \hline
                $()$ & $3$ & $0$ \\
                \hline
            \end{tabular}};
            \node[node] (JoinX3) at (3,5.8) {\begin{tabular}{|c|c|c|}
                \hline
                $V_{x_3}^{\Join}(x_3)$ & $w(t)$ & $\texttt{start}(t)$ \\
                \hline
                $d_1$ & $1$ & $0$ \\
                $d_2$ & $1$ & $1$ \\
                $d_3$ & $1$ & $2$ \\
                \hline
            \end{tabular}};
            \node[node, black!50] (ProjT) at (3,4.2) {\begin{tabular}{|c|}
                \hline
                $V_{(x_3,T)}^{\Pi}(x_3)$ \\
                \hline
                $d_1$ \\
                $d_2$ \\
                $d_3$ \\
                \hline
            \end{tabular}};
            \node[node, black!50] (JoinT) at (3,2.4) {\begin{tabular}{|c|}
                \hline
                $V_{T}^{\Join}(x_3)$ \\
                \hline
                $d_1$ \\
                $d_2$ \\
                $d_3$ \\
                \hline
            \end{tabular}};
            \node[node, black!50] (ProjX2) at (-3,4.4) {\begin{tabular}{|c|}
                \hline
                $V_{(x_1,x_2)}^{\Pi}(x_1)$ \\
                \hline
                $b_1$ \\
                $b_3$ \\
                \hline
            \end{tabular}};
            \node[node, black!50] (JoinX2) at (-3,2.9) {\begin{tabular}{|cc|}
                \hline
                \multicolumn{2}{|c|}{$V_{x_2}^{\Join}(x_1, x_2)$} \\
                \hline
                $b_1$ & $c_1$ \\
                $b_1$ & $c_2$ \\
                $b_3$ & $c_1$ \\
                \hline
            \end{tabular}};
            \node[node, black!50] (ProjR) at (-3,1.2) {\begin{tabular}{|cc|}
                \hline
                \multicolumn{2}{|c|}{$V_{(x_2,R)}^{\Pi}(x_1, x_2)$} \\
                \hline
                $b_1$ & $c_1$ \\
                $b_1$ & $c_2$ \\
                $b_3$ & $c_1$ \\
                \hline
            \end{tabular}};
            \node[node, black!50] (JoinR) at (-3,-0.5) {\begin{tabular}{|cc|}
                \hline
                \multicolumn{2}{|c|}{$V_{R}^{\Join}(x_1, x_2)$} \\
                \hline
                $b_1$ & $c_1$ \\
                $b_1$ & $c_2$ \\
                $b_3$ & $c_1$ \\
                \hline
            \end{tabular}};

            \draw[->, thick] (JoinVROOT.south) -- (ProjX1.north);
            \draw[->, thick] (JoinVROOT.south) -- (ProjX3.north);
            \draw[->, thick] (ProjX3.south) -- (JoinX3.north);
            \draw[->, black!50, thick] (JoinX3.south) -- (ProjT.north);
            \draw[->, black!50, thick] (ProjT.south) -- (JoinT.north);
            \draw[->, thick] (ProjX1.south) -- (JoinX1.north);
            \draw[->,  black!50, thick] (JoinX1.south) -- (ProjX2.north);
            \draw[->, black!50, thick] (ProjX2.south) -- (JoinX2.north);
            \draw[->, black!50, thick] (JoinX2.south) -- (ProjR.north);
            \draw[->, black!50, thick] (ProjR.south) -- (JoinR.north);
        \end{tikzpicture}
        \subcaption{View-tree for $V(T_{Q_2})$}
    \end{subfigure}

    \caption{Representation of q-tree, view-tree and query result for $Q_2(D)$ (see \Cref{exa:q2})}
    \label{fig:example_datastructure_weigth}
\end{figure}

    Note that, for $t \in Z_{\vroot}^{\Join}$, the value $w(t)$ is the total number of tuples in $Q_2(D)$.
\end{example}

To understand how weights and prefix sums are maintained after updates, we describe the implementation of a bucket $b$. We use the $B^+$-tree described in \Cref{sec:data_structure}, where the tuples of $b$ are sorted lexicographically by their values. Each leaf stores a bounded number of consecutive tuples $t$ of $b$, together with their weights $w(t)$. Each internal node $n$ stores the sum of the values stored in its children, that is, the total weight of the tuples stored in the leaves below $n$. Inserting or removing a tuple in $b$, or changing its weight, takes time $O(\log |D|)$, including the update of the sums stored in the internal nodes along the path from the leaf to the root. Propagating an update of the database to the $O(|Q|)$ views of the grounded-tree thus takes time $O(|Q| \log |D|)$.

The prefix sums are not stored: the value $\texttt{start}_b(t)$ of a tuple $t$ is computed when needed. To compute it, we traverse the $B^+$-tree from the root to the leaf of $t$. At each internal node, we add the values stored in the children located to the left of the child we visit, and, in the leaf, we add the weights of the tuples preceding $t$. Since the height is $O(\log |D|)$ and each node has a bounded number of children, computing the prefix sum of a tuple takes time $O(\log |D|)$.
Then, in a bucket $b$, given a value $k$, we can find in time $O(\log |D|)$ the last tuple $t$ sorted according to $f_b$ such that $\texttt{start}_b(t) < k$.

\begin{theorem}
    There exists an algorithm that, given a q-hierarchical SJFCQ $Q$, a hierarchy-consistent total order $\prec$ on its free variables, a q-tree $T_Q$ for $Q$ compatible with $\prec$ and a database instance $D$, answers ranked direct access for $Q(D)$ in the lexicographic order defined by $\prec$ with $O(|D| \cdot |Q|)$ preprocessing, $O(|Q| \log |D|)$ update time and $O(|Q| \log |D|)$ access time.
\end{theorem}

\begin{proof}
    This access algorithm is Algorithm 1: Lexicographic Direct-Access from \cite{Carmeli2023}, whose data complexity is $O(\log(|D|))$. This algorithm consists of getting the right bucket for each view of the view-tree, which is done in constant time as views are implemented as look-up tables, and then picking the correct tuple in those buckets, which takes $O(\log (|D|))$, as proven in \cite[Lemma\ 3.8]{Carmeli2023}. In combined complexity, as we look for this tuple for each variable of the order $\prec$, and we traverse each view of the view-tree, the complexity of the access algorithm is $O(|Q| \log(|D|))$. 
    Hence, combining this complexity with \Cref{prop:linearPreproc} and \Cref{lem:logUpdate-HC}, there exists an algorithm that answers ranked direct access for $Q(D)$ and $\prec$ with $O(|D|\cdot  |Q|)$ preprocessing, $O(|Q| \log |D|)$ update time and $O(|Q| \log |D|)$ access time.
\end{proof}

\section{Non-Hierarchy-Consistent Orders}\label{sec:NonHC}

In this section, given a q-hierarchical SJFCQ $Q$, an order $\prec$ over $\free(Q)$ and a database instance $D$, our goal is to develop an algorithm with $O(|D| \cdot |Q|)$ preprocessing and $O(|Q| \log |D|)$ update time for ranked direct access under updates (resp.\ ranked enumeration under updates) according to $\prec$ with $O(|Q| \log |D|)$ access time (resp.\ delay), i.e., linear preprocessing, logarithmic update, and logarithmic access (resp.\ delay) times in data complexity.

We will show that for ranked direct access under updates, such an algorithm conditionally does not exist.

\subsection{Ranked Direct Access Under Updates}

Our contribution on ranked direct access under updates relies on \Cref{thm:UnrankedEnumlb}. The following lower bound states that efficient ranked direct access under updates cannot be achieved for a q-hierarchical SJFCQ when the given order on its free variables is allowed to be non-hierarchy-consistent. Our lower bound holds in data complexity and applies to any q-hierarchical SJFCQ and non-hierarchy-consistent order on its free variables.

\begin{theorem}\label{thm:RankedDAlb}
    Fix a number $\varepsilon > 0$, $\varepsilon' > 0$, a q-hierarchical SJFCQ $Q$ and an order $\prec$ over $\free(Q)$. If $\prec$ is not hierarchy-consistent for $Q$, then there is no algorithm with polynomial preprocessing time and $O(|D|^{1 - \varepsilon})$ update time that, given a database instance $D$, answers ranked direct access for $Q(D)$ and $\prec$ with $O(|D|^{1 - \varepsilon'})$ access time, assuming the OMv-conjecture in data complexity.
\end{theorem}

To prove this lower bound for ranked direct access, we first show that an efficient ranked direct access algorithm for any non-hierarchy-consistent order on any q-hierarchical SJFCQ $Q$ would imply efficient enumeration under updates, for some non-q-hierarchical SJFCQ $Q'$, which according to \Cref{thm:UnrankedEnumlb} is not possible (even without an order) when assuming the OMv-conjecture.

\begin{lemma}\label{lem:RankedDAlb}
    Suppose there is a q-hierarchical SJFCQ $Q$ and a non-hierarchy-consistent order $\prec$ over $\free(Q)$, such that there exists an algorithm in $\langle f_p, f_a, f_u \rangle$ where $f_p$ is polynomial, $f_a$ and $f_u$ are $O(|D|^{1-\varepsilon})$ for some $\varepsilon$, and that, given a database instance $D$, answers ranked direct access for $Q(D)$ and $\prec$. Then there exists a non-q-hierarchical SJFCQ $Q'$, such that there is an algorithm in $\langle f_p, f_d, f_u \rangle$ that, given a database instance $D$, enumerates $Q'(D)$ with $f_d = O(f_a \log |D|)$.
\end{lemma}

To show this lower bound, we reduce the enumeration of a projection $Q'$ of $Q$ to ranked direct access under updates on $Q$, where $Q$ is a q-hierarchical SJFCQ with a non-hierarchy-consistent order over its variables. 
We choose the free variables of $Q'$ so that $Q'$ is not q-hierarchical. Given a database instance $D$, we enumerate the tuples of $Q'(D)$ as follows: we use ranked direct access on
$Q(D)$ to retrieve the $k\ith$ answer, and then find by binary search the smallest $k' > k$ such that the $k'$ answer has a different projection on the free variables of $Q'$. 
Each such answer is a new tuple of $Q'(D)$, which we output. Let us give the
formal proof of \Cref{lem:RankedDAlb}:

\begin{proof}[Proof of \Cref{lem:RankedDAlb}]
    Let $D$ be a database instance, $Q$ be a q-hierarchical SJFCQ, and $\prec$ be a non-hierarchy-consistent order over $\free(Q)$.
    Let $(x_1, \ldots, x_l)$ be the free variables of $Q$ listed in the order of $\prec$. Then there exist two variables $x_i, x_j \in \free(Q)$ such that $\atoms_Q(x_i) \subsetneq \atoms_Q(x_j)$ and $x_i \prec x_j$. Let $Q' \coloneq \Pi_{X_i}(Q)$ be the projection of $Q$ over ${X_i} \coloneq \{x_1, \ldots, x_i\}$.
    Thus, $\free(Q') = \free(Q) \setminus \{x_{i+1}, \ldots, x_l\}$, with $x_j \in \{x_{i+1}, \ldots, x_l\}$. Therefore, $Q'$ is not q-hierarchical, since condition (ii) of \Cref{def:q-hierarchical} is violated on variables $x_i$ and $x_j$. Indeed, $\atoms_{Q'}(x_i) \subsetneq \atoms_{Q'}(x_j)$ and $x_i \in  \free(Q')$, whereas $x_j \notin \free(Q')$.

    Suppose there exists an algorithm with update time $t_u$ that supports ranked direct access in time $t_a$ for $Q(D)$ and the order $\prec$. Let $N \coloneq |Q(D)|$ denote the number of answers of $Q$ on $D$.
    Since $Q$ is a q-hierarchical SJFCQ, we can retrieve $N$ in $O(1)$ data complexity, after a $O(|D|)$ preprocessing, and maintain this value in $O(\log |D|)$ update time \cite[Theorem 3.9]{Keppeler2020Answering}.
    Let $\precbar$ be the lexicographic order of $Q(D)$ induced by $\prec$, and let $(t_1, \ldots, t_N)$ be the answers of $Q(D)$ listed according to $\precbar$, where each $t_k \in (t_1, \ldots, t_N)$ can be computed in time $t_a$ using the algorithm assumed in the lemma statement.
    
    We now show how to design an efficient enumeration algorithm for $Q'(D)$. We start by checking if $N = 0$ (or if accessing $t_1$ returns \texttt{out-of-bounds}), in which case we output \texttt{end-of-enumeration} and terminate; otherwise we compute $t_1$ and output the tuple $s_1 \coloneq t_1[X_i]$; this step takes time $t_a$. Then, at each step, we compute the next tuple in the enumeration from the last outputted one. 
    To do this, suppose we have already output the $p\ith$ tuple $s_{p} \coloneq t_k[X_i]$ for some $k \in [N]$. We define the \emph{next relevant rank} $k' \coloneq \min \{ \ell \in [k, N] \mid t_k[X_i] \neq t_\ell[X_i] \}$ as the next index of an answer of $Q$ whose projection to $X_i$ changes relative to $t_k$. If no such $\ell$ exists, the enumeration algorithm will output the special value \texttt{end-of-enumeration} and terminate. Otherwise, the $(p+1)\ith$ tuple of the enumeration will be set as $s_{p+1} \coloneq t_{k'}[X_i]$.

    All that remains is to explain how we can efficiently find $s_{p+1}$ or detect termination. We first access $t_N$ and check whether $t_N[X_i] = s_p$; if so, the enumeration is finished, and the algorithm will output \texttt{end-of-enumeration}. Otherwise, we show how to compute the next relevant rank $k'$. To do this, we compute $k'$ by performing a binary search on the interval $[l,r] = [k, N]$. 
    At each step, let $m = \lfloor (l+r)/2 \rfloor$ and access the tuple $t_m$. If $t_k[X_i] \precbar t_m[X_i]$, then $k' \in [l, m]$, and we set $r \coloneq m$. Otherwise, we have $k' \in [m+1, r]$, and we set $l \coloneq m+1$.
    We repeat this process, updating the interval $[l,r]$ at each step, until
    we have $l = r$. At that point, one can easily see that we must have $k' = l$. This binary search requires $O(\log N)$ access calls, and therefore runs in time $O(t_a\log N)$.
    
    Thus, we obtain an algorithm with update time $t_u$ and delay $t_d = O(t_a\log N)$ for enumerating $Q'(D)$. Moreover, we have $N = |Q(D)| \leq |D|^{|Q|}$, therefore, we obtain delay $t_d = O(t_a \log |D|)$ in data complexity.
\end{proof}

Having proved our auxiliary result (\Cref{lem:RankedDAlb}), we are ready to conclude the proof of \Cref{thm:RankedDAlb}:

\begin{proof}[Proof of \Cref{thm:RankedDAlb}]
    Let $D$ be a database instance, $Q$ be a q-hierarchical SJFCQ, and $\prec$ be a non-hierarchy-consistent order over $\free(Q)$ for $Q$. Suppose there is $\varepsilon > 0$ such that there exists an algorithm with polynomial preprocessing and $O(|D|^{1 - \varepsilon})$ update time that answers ranked direct access in $O(|D|^{1-\varepsilon})$ access time for $Q(D)$ and $\prec$.
    
    By \Cref{lem:RankedDAlb}, there exists a non-q-hierarchical SJFCQ $Q'$ such that there is an algorithm that enumerates $Q'(D)$ with $O(|D|^{1-\varepsilon} \log |D|)$ delay and sublinear update time. 
    Let $\varepsilon' \coloneq \varepsilon/2$. By \Cref{thm:UnrankedEnumlb}, as $Q'$ is not q-hierarchical, no algorithm enumerates $Q'(D)$ with $O(|D|^{1 - \varepsilon'})$ delay and $O(|D|^{1 - \varepsilon'})$ update time, unless the OMv-conjecture fails. Since $O(|D|^{1 - \varepsilon} \log |D|) = O(|D|^{1 - \varepsilon'})$ this contradicts the OMv-conjecture.
    
    Therefore, unless the OMv-conjecture fails, there is no dynamic algorithm with polynomial preprocessing time and $O(|D|^{1-\varepsilon})$ update time that answers ranked direct access in $O(|D|^{1-\varepsilon'})$ access time for a q-hierarchical SJFCQ and a non-hierarchy-consistent order on its free variables.
\end{proof}

\subsection{Ranked Enumeration Under Updates}

We have shown a dichotomy for ranked direct access under updates: assuming
the OMv-conjecture, there exists an algorithm for ranked direct access with
complexity $\langle |D| \cdot |Q|, |Q| \log |D|, |Q| \log |D| \rangle$ if and
only if the order in input is hierarchy-consistent. For ranked enumeration, we
do not have a matching lower bound. We believe in fact that the situation
with ranked enumeration under updates is different, and that it remains efficient even when the order is not hierarchy-consistent, at the cost of an additional logarithmic factor in the delay in data complexity.

Our approach is to adapt the algorithm anyK-part of \cite{tziavelis2025any}, which relies on the Lawler-Murty procedure, to the grounded tree $V_D(T_Q)$. The main difference with the static setting lies in the optimal weights of the tuples. In the static case, they are computed once, before the enumeration, by a dynamic programming pass over the tree. In our setting, we compute them during the preprocessing phase, as part of the construction of the grounded tree, and we maintain them when the database is updated. We leave the details and the analysis of this approach to a future version of the present work.

\bibliographystyle{alpha}
\bibliography{biblio}

\end{document}